\documentclass{appolb}
\usepackage{graphicx}

\begin{document}
\title{Heavy-flavor jet properties and correlations from small to large systems with ALICE%
\thanks{Presented at Quark Matter Conference}%
}
\author{Marianna Mazzilli for the ALICE Collaboration
\address{CERN}
}
\maketitle
\begin{abstract}
The early production of heavy-flavor (HF, charm and beauty) quarks makes them an excellent probe of the dynamical evolution of quantum chromodynamics (QCD) systems. Jets tagged by the presence of a HF hadron give access to the kinematics of the heavy quarks, and along with correlation measurements involving HF hadrons allow for comparisons of their production, propagation and fragmentation across different systems. In this contribution the latest results on HF jets and correlations measured with the ALICE detector in pp, p--Pb and Pb--Pb collisions from the LHC Run 2 data are reported.

\end{abstract}
  
\section{Introduction}
HF jets and correlations can provide additional information with respect to single-hadron studies since they allow one to access the original parton kinematics. In pp collisions, the measurement of HF jet production is largely sensitive to the heavy-quark (HQ) production processes (leading order or next-to-leading order), which is useful to test pQCD calculations. 
The angular evolution of the jet parton shower depends on the mass of the initiating parton through a phenomenon known as dead-cone effect. Eventually the shower evolves into a multi-particle final state of experimentally detectable hadrons. Tagging jets by the presence of HF hadrons allows one to further characterize the hadronization mechanism. In larger systems, HF jets and correlations can be used to explore cold-nuclear matter effects in p--Pb collisions and to probe the microscopic interactions of the parton with the quark--gluon plasma (QGP) in Pb--Pb collisions.

\section{Heavy flavor jets and correlations in pp collisions}
\subsection{Heavy quark production}
\begin{figure}[h!]
    \centering
    \includegraphics[width=0.8\textwidth]{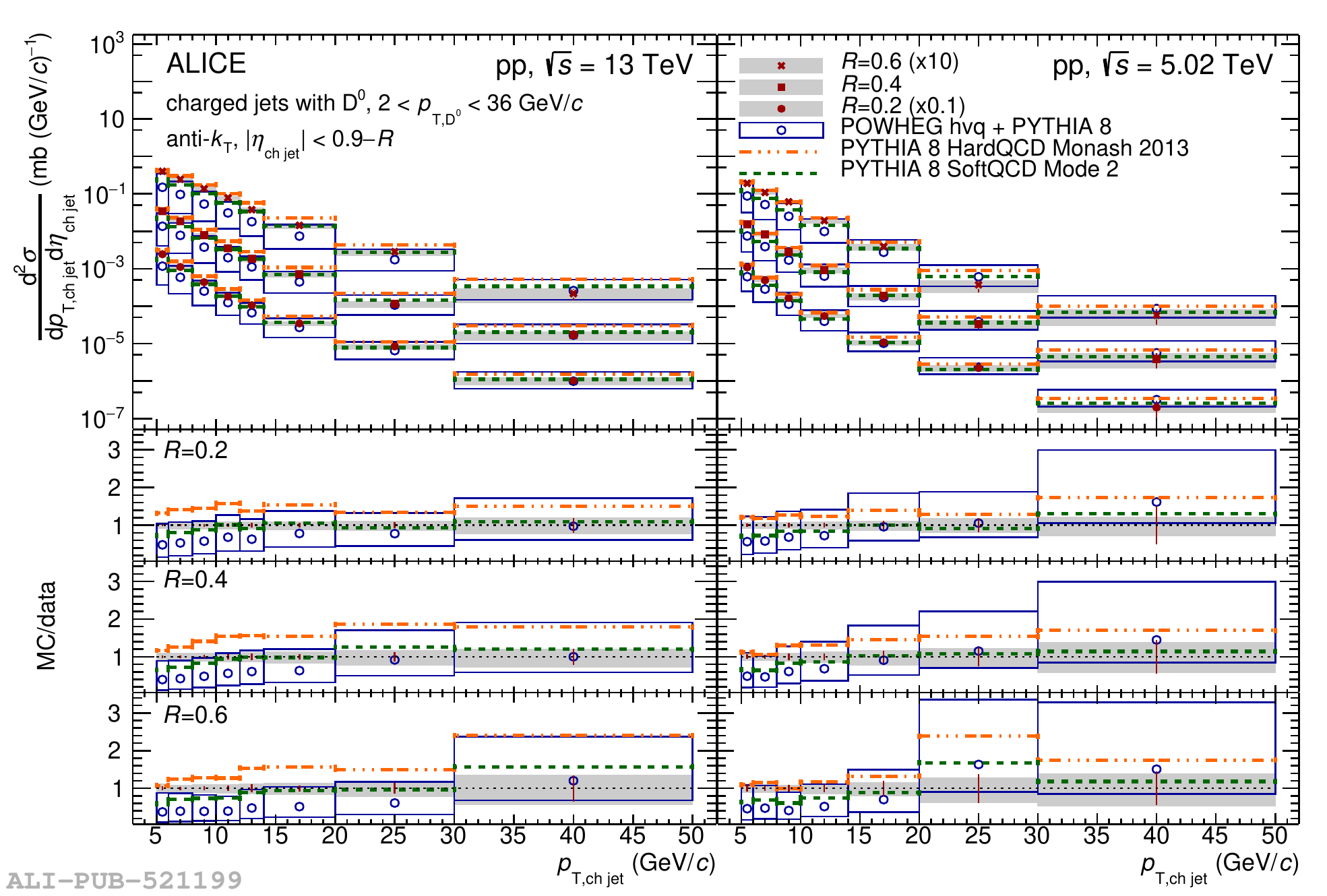}
    \caption{Top panels: $p_{\mathrm{T}}^{\mathrm{ch~jet}}$-differential cross section of charm jets tagged with D$^{0}$ mesons for $R$ = 0.2, 0.4, 0.6 in pp collisions at $\sqrt{s}$~=~13 TeV (left) and 5.02 TeV (right) compared to PYTHIA 8 HardQCD Monash 2013, PYTHIA 8 SoftQCD Mode 2, and POWHEG hvq + PYTHIA 8 predictions. Bottom panels: ratios of MC predictions to the data for $R$ = 0.2, 0.4 and 0.6.}
    \label{fig:xsec}
\end{figure}

The production of charm and beauty jets in pp collisions at $\sqrt{s}$~=~5.02~TeV and 13 TeV was measured with ALICE~\cite{ALICE:2022mur, ALICE:2021wct}. Charm jets were identified by the presence of a prompt D$^{0}$ meson (reconstructed via its hadronic decay $\mathrm{D^{0}}\to \mathrm{K}\pi^{-}$) among their constituents, while beauty jets were tagged exploiting the wider impact parameter distribution of beauty-hadron-decay particles. Charm (beauty) charged-jet reconstruction was performed using the track-based procedure with the FastJet~\cite{Cacciari:2011ma} anti-k$_\mathrm{T}$ algorithm~\cite{Cacciari:2008gp} with resolution parameters $R$~=~0.2, 0.4 and 0.6 ($R$~=~0.4), in the kinematic range 5$~<p_{\mathrm{T}}^{\mathrm{ch~jet}}<~$50 GeV/$c$ (10$~<p_{\mathrm{T}}^{\mathrm{ch~jet}}<~$100 GeV/$c$). 
In Fig.~\ref{fig:xsec}, the $p_{\mathrm{T}}^{\mathrm{ch~jet}}$-differential cross sections of D$^{0}$-tagged jets in pp collisions with resolution parameter $R$~=~0.2, 0.4 and 0.6 at $\sqrt{s}$~=~5.02 TeV and 13 TeV are reported. The results are compared to Monash-2013~\cite{Sjostrand:2014zea} and Mode 2~\cite{Christiansen:2015yqa} tunes of PYTHIA 8 with HardQCD and SoftQCD processes, respectively, and with POWHEG + PYTHIA8~\cite{Frixione:2007vw,Nason:2004rx}. The PYTHIA 8 predictions with the SoftQCD and Mode 2 tune settings provide the best description of the $p_{\mathrm{T}}^{\mathrm{ch~jet}}$-differential cross sections for both collision energies and all resolution parameters. Within the experimental and theoretical uncertainties, the measurements are also in agreement with the POWHEG + PYTHIA 8 calculations. The $p_{\mathrm{T}}^{\mathrm{ch~jet}}$-differential inclusive production cross section of b jets, as well as the corresponding inclusive b-jet fraction, are reported in~\cite{ALICE:2021wct} and the measurements are well reproduced by POWHEG calculations with PYTHIA8 fragmentation.


\subsection{Looking into the parton shower}
\begin{figure}[h!]
    \centering
    \includegraphics[width=1.0\textwidth]{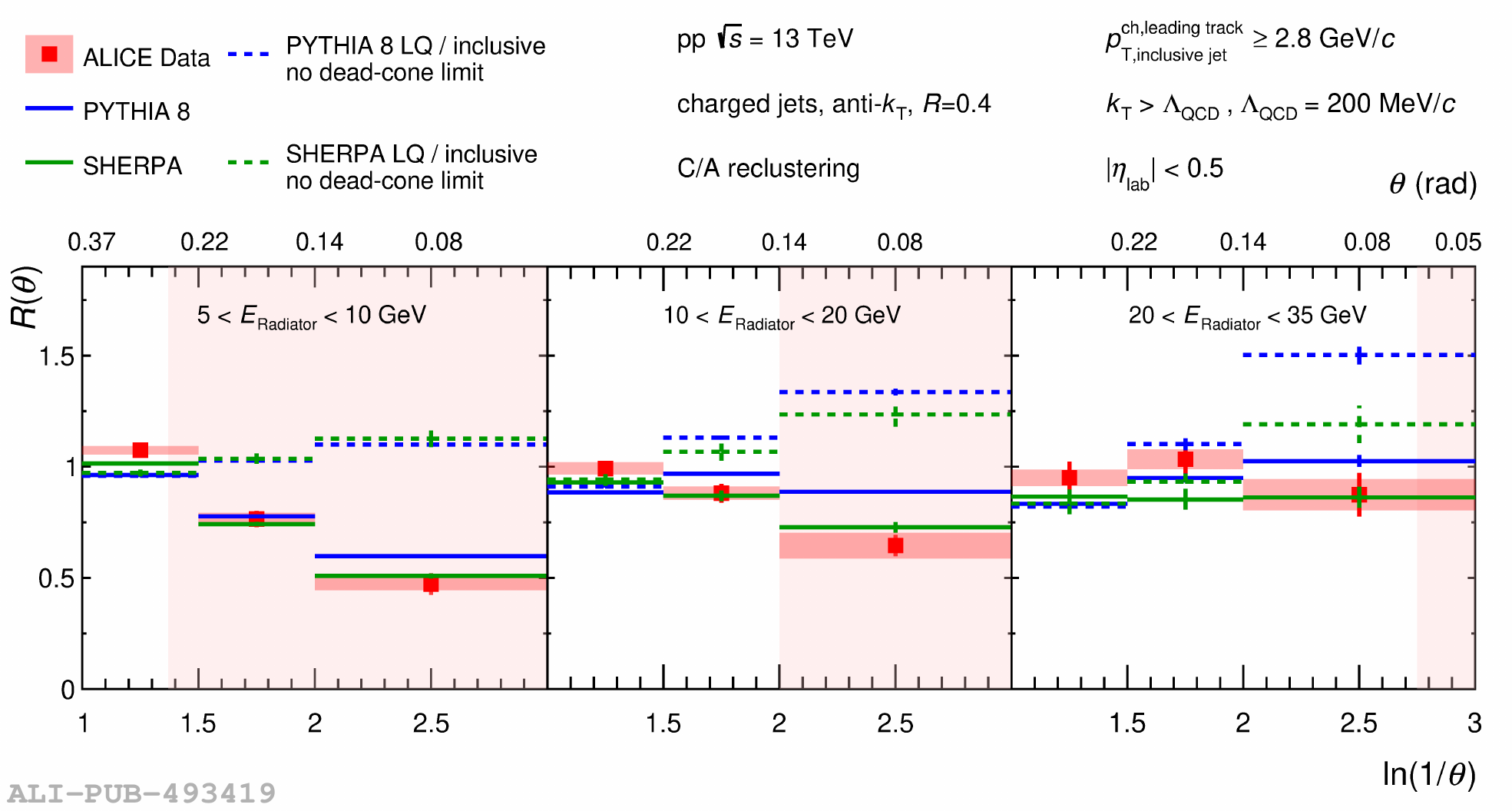}
    \caption{Ratios of the splitting-angle probability distributions for D$^{0}$-meson tagged jets to inclusive jets, $R$($\theta$) in pp collisions at $\sqrt{s}$ = 13 TeV for 5~$<~E_\mathrm{Radiator} <$~10~GeV, 10 $< E_\mathrm{Radiator} <$~20 GeV and 20~$<E_\mathrm{Radiator}~<$~35~GeV. The data are compared with PYTHIA 8 and SHERPA predictions.}
    \label{fig:deadcone}
\end{figure}

The pattern of the parton shower is one of the key experimental tools in understanding the properties of QCD~\cite{Tanabashi:2018oca}. This pattern depends on the mass of the initiating parton through the dead-cone effect, which predicts a suppression of the gluon spectrum emitted by a HQ of mass $m$ and energy $E$, within a cone of angular size $\theta\approx m/E$ around the emitter. The observable used to reveal the dead-cone effect is built by constructing the ratio $R(\theta)$ of the splitting angle distributions for D$^{0}$-tagged jets and inclusive jets in intervals of the energy of the radiator~\cite{ALICE:2021aqk}. 
In Fig.~\ref{fig:deadcone} the measured $R(\theta)$ is reported in three different energy intervals and it shows a significant suppression at small-splitting angles for D$^{0}$-tagged jets with respect to inclusive jets. The data are compared with PYTHIA 8~\cite{Sjostrand:2014zea} and SHERPA~\cite{Sherpa:2019gpd} simulations, including the no dead-cone limit given by the ratio of the angular distributions for light-quark jets (LQ) to inclusive jets. In the limit where the dead-cone is not present, the ratio is larger than unity. The shaded areas correspond to the angles within which emissions are suppressed by the dead-cone effect, assuming a charm-quark mass of 1.275 GeV/$c^2$. The magnitude of the suppression increases with decreasing radiator energy.

\subsection{Heavy flavor jets and correlations to look into fragmentation}
A way to directly probe the fragmentation of quarks into hadrons is to describe the hadron momentum in relation to the momentum of the jets using the following observable: 
\begin{equation}
    z_{\parallel}^\mathrm{ch}=\frac{\vec{p}_\mathrm{D^0}\cdot\vec{p}_\mathrm{ch~jet}}{\vec{p}_\mathrm{ch~jet}\cdot\vec{p}_\mathrm{ch~jet}}.
\end{equation}
The $z_{\parallel}^\mathrm{ch}$ distributions was measured for D$^{0}$-meson tagged jets in pp collisions at $\sqrt{s} = 5.02$ TeV and 13 TeV~\cite{ALICE:2022mur}. At low transverse momentum of the jet and larger radii, the momentum fraction carried by the D$^{0}$ mesons shows a hint of softer fragmentation in data when compared to POWHEG and PYTHIA8 model predictions. 

\begin{figure}[h!]
    \centering
   \includegraphics[width=0.85\textwidth]{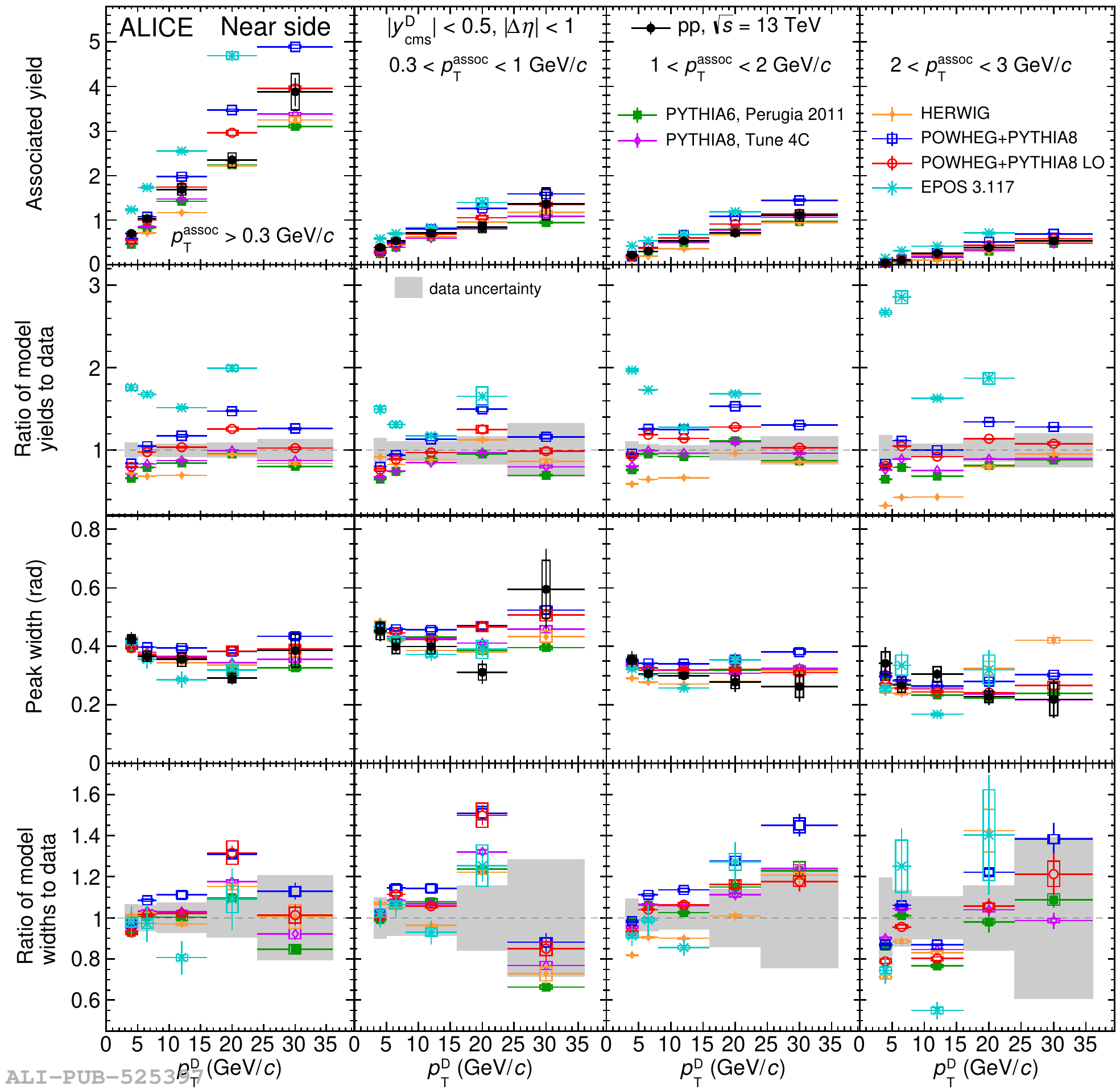}
   \caption{Near-side associated peak yields (top row) and widths (third row down) in pp collisions at $\sqrt{s}$ = 13 TeV, compared with predictions from PYTHIA6 and PYTHIA8, POWHEG+PYTHIA8, HERWIG, and EPOS 3 event generators with various configurations. The ratios of model predictions to data measurements for yield (width) values are shown in the second (fourth) row.}
    \label{fig:corr}
\end{figure}

An alternative approach to probe the fragmentation is to study the properties of the near-side peak of azimuthal-angle correlations of D mesons with charged particles, which are strongly connected to the charm jet properties in terms of particle multiplicity and angular profile~\cite{ALICE:2019oyn, ALICE:2021kpy}. In Fig.~\ref{fig:corr} the measured near-side peak yields and widths were compared with expectations from PYTHIA6 and PYTHIA8, POWHEG+PYTHIA8, HERWIG~\cite{Bahr:2008pv}, and EPOS 3~\cite{Werner:2010aa} models. Among these, PYTHIA8 and POWHEG+PYTHIA8 provide the best overall description of the data.

\section{Heavy flavor jets in p--Pb collisions}
The beauty-jet production cross section has been measured down to $p_\mathrm{T}^\mathrm{ch~jet}$=10 GeV/$c$ in p--Pb collisions~\cite{ALICE:2021wct}. The overall impact of cold nuclear matter effects on the resulting ${p}_\mathrm{T}^\mathrm{ch~jet}$-differential cross section can be quantified by means of the nuclear modification factor $R_{\mathrm{pPb}}^\mathrm{b-jet}$ defined as the ratio of the yield measured in p--Pb collisions and the expected yield that would be obtained from a superposition of independent pp collisions. The $R_{\mathrm{pPb}}^\mathrm{b-jet}$ for beauty jets, reported in~\cite{ALICE:2021wct}, is compatible with unity and does not show evidence of cold nuclear matter effects on beauty jets in p--Pb collisions in the measured kinematic range within the current precision. The measurements are well reproduced by POWHEG NLO pQCD calculations with PYTHIA8 fragmentation.

\section{Heavy flavor jets in Pb--Pb collisions}
\begin{figure}[h!]
    \centering
    \includegraphics[width=0.48\textwidth]{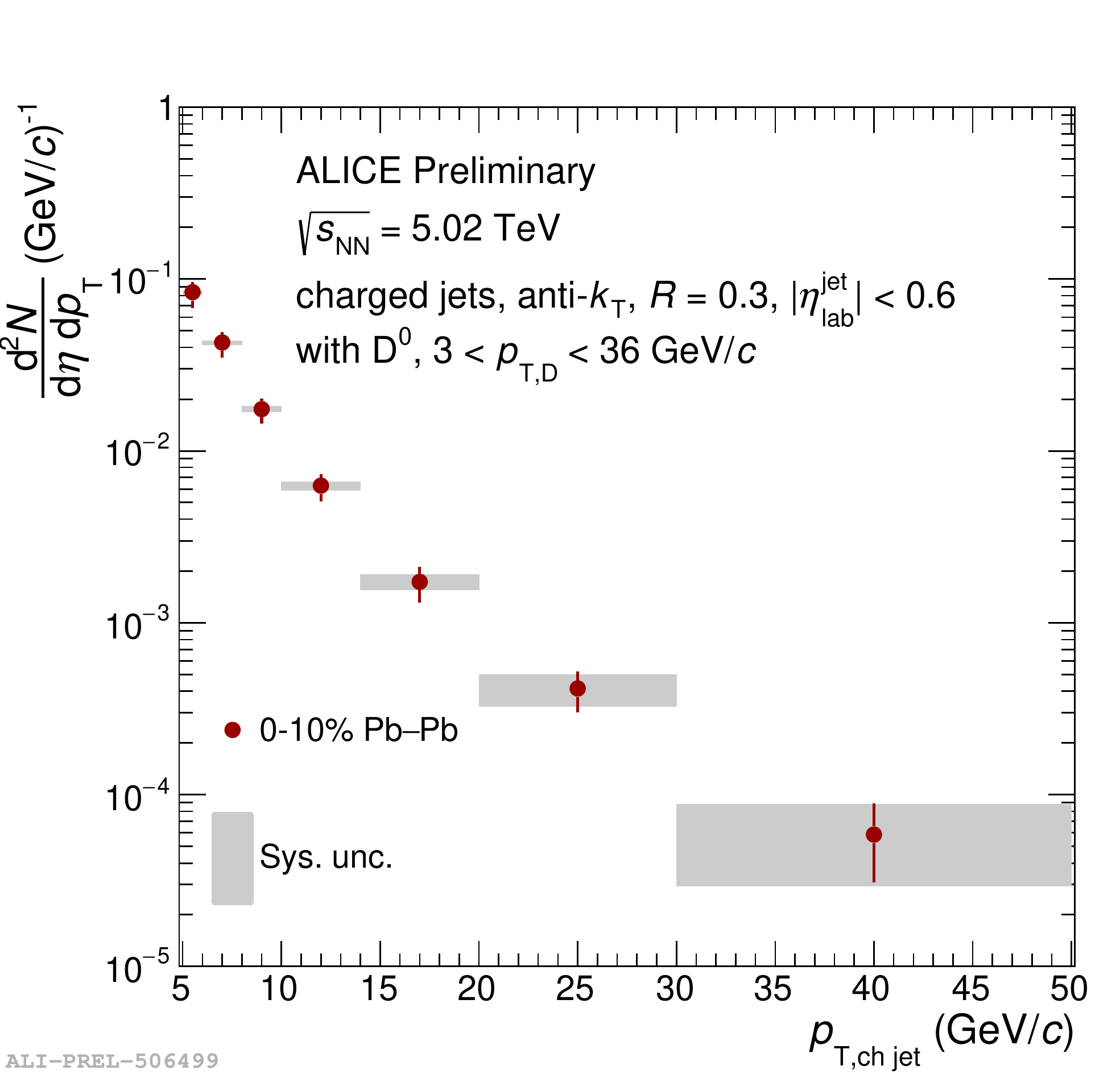}
    \includegraphics[width=0.48\textwidth]{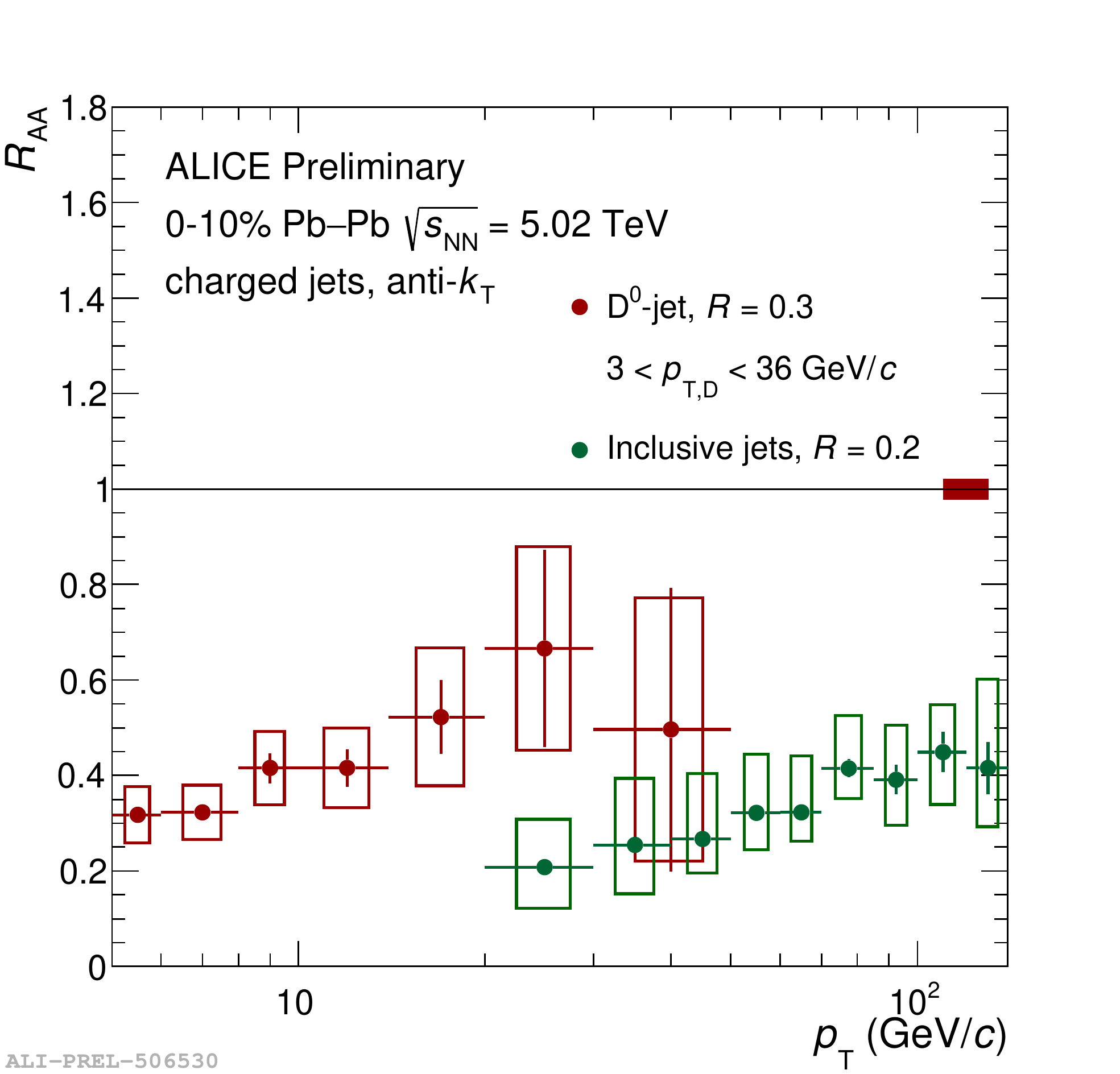}
    \caption{Left panel: $p_{\mathrm{T}}^{\mathrm{ch~jet}}$-differential cross section of charm jets tagged with D$^{0}$ mesons for $R$ = 0.3 in Pb--Pb collisions at $\sqrt{s_\mathrm{NN}}$~=~5.02 TeV. Right panel: nuclear modification factor $R_\mathrm{AA}$ of D$^{0}$-tagged jets with $R$ = 0.3 (red) and inclusive jets with $R$ = 0.2 (green) as a function of ${p}_\mathrm{T}^\mathrm{ch,jet}$.}
    \label{fig:Raa}
\end{figure}

The $p_\mathrm{T}$-differential yield of D$^{0}$-tagged jets with $R$~=~0.3 was measured for 5~$<~p_{\mathrm{T}}^{\mathrm{ch~jet}}<$~50 GeV/$c$ in central Pb--Pb collisions and it is reported in the left panel of Fig.~\ref{fig:Raa}. 
In order to quantify the modification of the jet spectrum in heavy-ion collisions due to the interactions with the QGP, the nuclear modification factor $R_{\mathrm{AA}}$ was measured and it is shown in the right panel of Fig.~\ref{fig:Raa}. A strong suppression is observed, with a hint of a lower $R_{\mathrm{AA}}$ compared to the inclusive charged jets with $R$ = 0.2~\footnote{The $R_{\mathrm{AA}}$ of inclusive jets was measured for $R$~=~0.2 and $R$~=~0.4 and it was found to be compatible, within uncertainties, between the two different resolution parameters, justifying the comparison to the $R_{\mathrm{AA}}$ of D$^{0}$-tagged jets with $R$~=~0.3. } measurement in the two  higher $p_\mathrm{T}^\mathrm{ch~jet}$ intervals measured for D$^{0}$-tagged jets. 

\bibliographystyle{unsrt} 
\bibliography{bibliography_old}




\end{document}